\documentclass[twocolumn,showpacs,preprintnumbers,amsmath,amssymb,10pt,a4paper]{revtex4}
\usepackage{graphicx}
\usepackage{epsfig}
\usepackage[latin1]{inputenc}
\usepackage{amsmath}
\usepackage{amsfonts}
\usepackage{amssymb}
\usepackage{dcolumn}
\usepackage{bm}

\begin{document}


\title{Static and dynamic properties of vortices in anisotropic magnetic disks}%

\author{Tiago S. Machado$^{1}$}
\author{Tatiana G. Rappoport$^{2}$}
\author{Luiz C. Sampaio$^{1}$}%

\affiliation{%
$^{1}$Centro Brasileiro de Pesquisas F\'{\i}sicas, Xavier Sigaud, 150, Rio de Janeiro, RJ, 22.290-180, Brazil
}%

\affiliation{%
$^{2}$Instituto de F\'{\i}sica, Universidade Federal do Rio de Janeiro, Rio de Janeiro, RJ, 68.528-970, Brazil
}%


\date{\today}

\begin{abstract}

We investigate the effect of the magnetic anisotropy ($K_z$) on the static and dynamic properties of magnetic vortices in small disks. Our micromagnetic calculations reveal that for a range of  $K_z$ there is an enlargement of the vortex core. We analyze the influence of $K_z$ on the dynamics of the vortex core magnetization reversal under the excitation of a pulsed field. The presence of $K_z$, which leads to better resolved vortex structures, allows us to discuss in more details the role played by the in-plane and perpendicular components of the gyrotropic field during the vortex-antivortex nucleation and annihilation.

\end{abstract}

\maketitle


The manipulation of magnetization in nanostructured materials by means of magnetic field and/or spin polarized current has attracted substantial attention in the last decade.  More recently, a special focus on the study of magnetization reversal dynamics in magnetic disks~\cite{Yamada2007} and lines~\cite{Parkin2008} has been motivated mainly by their potential importance in the implementation of memory and logical operations.

Micro-sized Permalloy (Py, FeNi) disks can exhibit, depending on their size and aspect ratio, a magnetic vortex with a core ($\sim10-20$ nm) magnetized perpendicular to the disk plane~\cite{Wachowiak2002}. Due to the magnetic bi-stability of the vortex structure, they have been considered for technological applications. The vortex core magnetization reversal can be achieved by applying an in-plane magnetic field or spin polarized current in the form of short pulses~\cite{Hertel2007} and/or alternating (AC) resonant  excitation~\cite{Yamada2007,Waeyenberge2006,Kim2007} - both have an equivalent role on determining the vortex core dynamics.

It is usually considered that the size of the vortex core depends on parameters such as exchange constant, thickness and diameter of the magnetic disk. Most of the research on vortices in magnetic systems neglects the effect of magnetic anisotropy. On the other hand, it has been demonstrated that a uniaxial magnetic anisotropy in Permalloy particles can be induced  by the deposition process~\cite{Eames2002}. Experiments and simulations have also shown that the presence of anisotropy in thick magnetic disks gives rise to a diversity of domain patterns~\cite{Eames2002,Moutafis2006}.

In this letter, we study the role played by the magnetic anisotropy in the magnetic properties of disks. Using micromagnetic simulations, we consider a magnetic anisotropy ($K_z$) perpendicular to the disk plane and analyze how it modifies the magnetization pattern and the dynamics of the vortex core under action of an in-plane pulsed magnetic field. We also discuss in detail the core reversal process and the influence of $K_z$ on the gyrotropic field and the magnetization reversal time.

The simulations were performed with a code we developed, which employs the Landau-Lifshitz-Gilbert (LLG) equation.  We used the typical magnetic parameters of Py: the saturation magnetization is given by $M_s=8.6 \times 10^5$ A/m, the exchange coupling is $A=1.3\times10^{-11}$ J/m and the Gilbert damping constant is $\alpha=0.2$. The magnetic anisotropy was included in the total effective magnetic field $\bm{h_{\rm eff}}$ and is given by $(2K_z/\mu_{0} M_s^2)m_{z}{{\hat{\rm z}}}$ \cite{Fidler}. $K_{z}$ varies from $0$ to $10^6$ J/m$^3$. We have simulated Py disks with the diameter of $300$ nm and thickness of $12$ nm. The disk is discretized in cells of $3\times 3 \times 3 $ nm$^3$.

\begin{figure}[h]
\includegraphics[width=0.7\columnwidth]{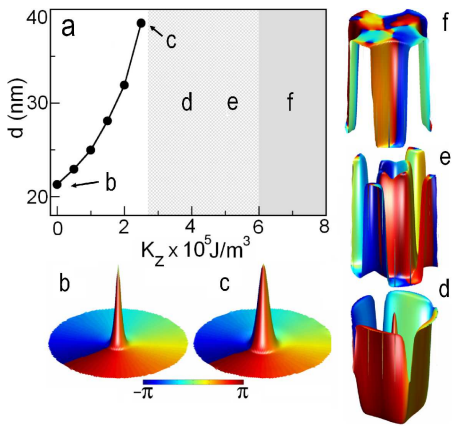}
\caption{(a) The diameter of the vortex core as a function of $K_z$ for a disk with diameter of 300 nm and thickness of 12 nm. Panels (b-f)  show the magnetization pattern for increasing values of $K_z$, illustrating  the different regimes of (a). The colors represent the direction of the in-plane component of the magnetization $m_{xy}$. A vortex structure is given by a clockwise color sequence of blue-green-red.\label{Fig1}}
\end{figure}

Let us first consider the vortex core structure in static equilibrium (in the absence of an external magnetic field).  The magnetization pattern of the Py disk presents three characteristic regimes, as can be seen in Fig.~\ref{Fig1} (b-f): for $K_z$ below $2.5 \times 10^5$ J/m$^3$ the main consequence of increasing the anisotropy is an increase of the vortex core diameter. The size of the vortex core is measured at half of the maximum value of $m_z$ and its dependence with $K_z$ is shown in Fig.~\ref{Fig1}(a). For $K_z$ between $2.5 \times 10^5$ and $4.0 \times 10^5$ J/m$^3$, $m_z$ exhibits  concentric regions with $+M_s$ and $-M_s$, still preserving the core in the center of the disk. For $K_z$ between $4 \times 10^5$ and $6 \times 10^5$ J/m$^3$ the core disappears and the number of concentric rings increases, which also happens for larger disks.  Moreover, in these two ranges, the in-plane magnetization still preserves the vorticity (see Figs.\ref{Fig1}d-e). The similarity between these patterns and previous observed patterns in Co nanomagnets~\cite{Hehn1996} and thick NiFe nanodisks~\cite{Eames2002} is noteworthy. In both cases, the concentric rings are clearly seen in the images obtained by Magnetic Force Microscopy (MFM). For $K_z$ above $6 \times 10^5$ J/m$^3$, the vortex structure is lost and we see what amount to a single domain in $m_z$ (Fig. \ref{Fig1}f). For larger and thinner disks,  the magnetization pattern is composed by stripes.

The prospect of having large vortex cores has clear advantages for magnetization detection. Still, in order to consider the practical aspects it is necessary to study the stability of these vortex cores and the possibility of switching their magnetization.  For this purpose,  we studied the dynamics of the system under short in-plane magnetic field pulses. We simulated the magnetic reversal process and constructed a switching diagram. Similar diagrams for pulse parameters have been constructed recently for magnetic disks~\cite{Hertel2007}.  They show that the operating field range is narrow:  low fields do not produce core switching. On the other hand, higher fields can give rise to multiple switchings. 
\begin{figure}[h]
\includegraphics[width=1.0\columnwidth]{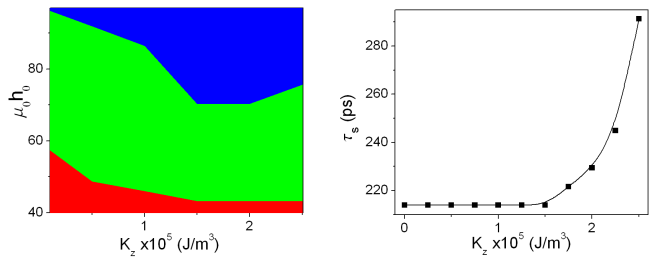}
\caption{(a) Switching vortex core magnetization diagram for magnetic field pulse strength and $K_z$. The pulse duration is fixed ($263$ ps). Red, green, and blue colors represent no, one, and multiple switches, respectively. In the grey area the vortex core is expelled from the disk. (b) The switching time as a function of $K_z$ for a fixed field strength $B_0=64$ mT.\label{Fig2}}
\end{figure}

With the intention of discussing the influence of $K_z$ on this process, we built the diagram sketched in Fig. {\ref{Fig2}a}. We use a pulse of Gaussian form with a fixed pulse duration ($263$ ps) and a variable field strength $B_0$.  For each value of $K_z$,  we count the number of core magnetization inversions  during a single pulse length. Fig. {\ref{Fig2}a} shows three different dynamical regimes  in response to the exciting field. In the absence of magnetic anisotropy, the field strength that is necessary to switch the core magnetization is $B_0=60$ mT. However, for fields higher than 95 mT undesirable multiple switches are produced.  For disks with $K_z \neq 0$, the pulse necessary to induce the switching process has a strength comparable to the $K_z=0$ case.  As can be seen in Fig. {\ref{Fig2}a}, an important consequence of increasing $K_z$ is the decrease of the minimum field necessary to produce a single switch. Such dependence on $K_z$, for the minimal field for the switching, opens the possibility of producing selective vortex inversions in a group of magnetic disks with different $K_z$'s. Note that under the action of a pulsed field, the core moves towards the disk border in a curved (or spiraling) trajectory. When the pulse is over, the core returns back to the disk center. Depending on the pulse strength and duration, the vortex core can be expelled from the disk during this process. Such behavior is not shown in our diagram and can be avoided by increasing the disk diameter. 

Another interesting aspect of the influence of $K_z$ can be seen in the switching time $\tau_{s}$.  $\tau_{s}$ is the interval between the pulse start and the complete reversal of the vortex core magnetization. We have calculated $\tau_{s}$ for a given pulse strength ($B_0=64$ mT) in the single switching regime (see the horizontal line in the Fig.{\ref{Fig2}a}). $\tau_{s}$ for $K_z=0$ is $213$ ps and it increases monotonically with $K_z$, reaching  values $\simeq 1.4$ times larger than $\tau_{s} (K_z=0)$ (Fig.{\ref{Fig2}b}). For thinner disks ($6$nm) $\tau_s$ can reach up to $2.5 \tau_s (K_z=0)$.

\begin{figure}[h]
\includegraphics[width=0.7\columnwidth]{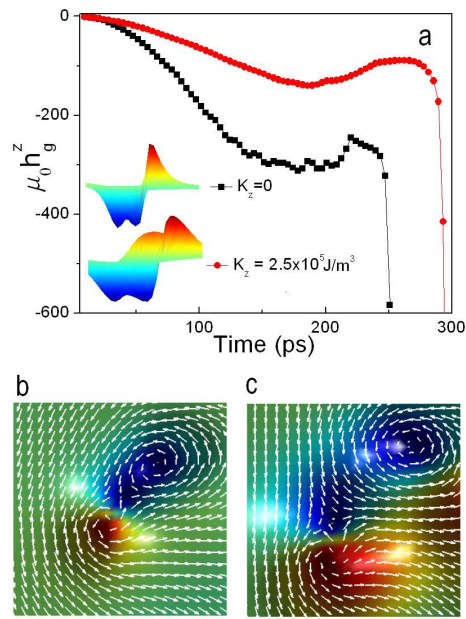}
\caption{(a) Time evolution of the gyrotropic field (perpendicular component). Snapshots of vortices and antivortex for (b) $K_z=0$  at the maxima of the gyrotropic field  and (c) $K_z=2.5\times10^5$ J/m$^3$ and $t=263$ ps. The color map represents the perpendicular component of the magnetization ($m_z$) and the arrows show the in-plane component. The inset of panel (a) gives a transverse view of $m_z$.\label{Fig3}}
\end{figure}

A closer look at the core reversal dynamics for disks with $K_z\neq 0$  shows that the intermediate processes leading to the switching are similar to the ones obtained in previous analysis for $K_z = 0$ ~\cite{Yamada2007,Waeyenberge2006,Hertel2007,Xiao2006,Kim2008}: the core shape changes during the magnetization reversal,  through the formation of the adjacent V-AV (vortex-antivortex) pair and the subsequent V-AV annihilation and nucleation of the reversed vortex. However, in opposition to what is normally observed in micromagnetic calculations, here,  due to the increase of the vortex core diameter, the formation of the adjacent V-AV pair during the reversal process can be well resolved (see inset of Fig.~\ref{Fig3} and Figs.~\ref{Fig3}(b) and (c)).

\begin{figure}[h]
\includegraphics[width=0.65\columnwidth]{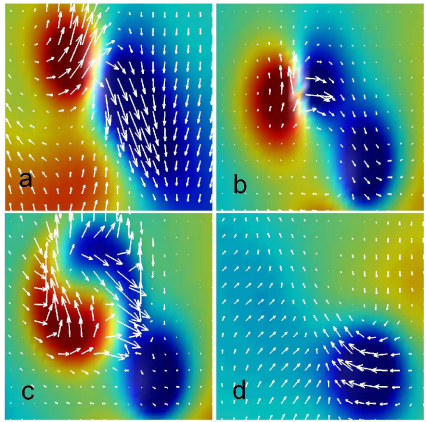}
\caption{In-plane component of the gyrotropic field (arrows) just before (a) and after (b-d) the nucleation and separation of the vortex and antivortex with negative magnetizations. The color map represents the out-of-plane magnetization.\label{Fig4}}
\end{figure}

The vortex core magnetization reversal, its switching time, the core motion and its deformations can be understood in terms of the gyrotropic field~\cite{Yamada2007,Kim2008}, which acts on the vortex core only during its movement. The gyrotropic field is given by $\textbf{h}_{g}=\frac{1}{\gamma} \textbf{m}\times [ \textbf{v}\cdot \nabla \textbf{m}]$\cite{Yamada2007,Kim2008},
where $\gamma$ and $\textbf{v}$ are the gyromagnetic factor and the core velocity, respectively. Indeed, using Thiele's equation~\cite{Thiele1973} in its original form,  $\textbf{h}_{g}$ can be simplified using $\dot{\textbf{m}}$ instead of $\textbf{v}$, and it is reduced to $\textbf{h}_{g}=\frac{1}{\gamma} \textbf{m} \times \dot{\textbf{m}}$. All calculations shown below were performed with this last expression of $\textbf{h}_{g}$.

Previous discussions on the gyrotropic field have emphasized the role of the perpendicular component of this effective field in the reversal process\cite{Kim2008}. To further investigate this, we have calculated the perpendicular $\textbf{h}_{g}^{z}$ and the in-plane $\textbf{h}_{g}^{xy}$ component as well. First, we discuss the time evolution of the perpendicular component $\textbf{h}_{g}^{z}$ for $K_z=0$ and $2.5\times10^5$J/m$^3$, which is shown  Fig.~\ref{Fig3}a. Following the application of the field pulse, the core moves and $\textbf{h}_{g}^{z}$ acts on the core at the side opposite to the movement direction leading to the formation of a peak with negative magnetization. At this stage, it is just a peak without a vortex structure. $\textbf{h}_{g}^{z}$ increases (in modulus) up to the point where the peak is so wide that it leads to the nucleation of a V-AV pair, labelled as V$^-$-AV$^-$, respectively. In conjunction with this  V$^-$-AV$^-$ nucleation there is a decrease (in modulus) of $\textbf{h}_{g}^{z}$. Subsequently, the  AV$^-$-V$^+$ annihilation takes place at the  $\textbf{h}_{g}^{z}$ divergence. This divergence is in agreement with Ref.~\onlinecite{Kim2008}. Such behavior is independent of $K_z$ but it is worth noticing that the separation of the V$^-$-AV$^-$ is better resolved spatially and  takes place after an interval that is longer than the one for $K_z=0$. This can be seen in the snapshots in the inset of Fig.~\ref{Fig3}, and in the comparison of Figs.~\ref{Fig3} (b) and (c). On the right one sees the V$^+$ with a close neighbor, at its left, which is the  AV$^-$. Further on the left, one sees the V$^-$. As can be gathered from the figures, $\textbf{h}_{g}^{z}$ decreases for increasing values of  $K_z $ and that is the reason for the increase of $\tau_{s}$.

One of the main advantages of considering $K_z\neq$ 0 in our calculations is the large separation between the Vs and AVs involved in the switching process. This allows us to analyze in more details  the structure of the gyrotropic field in between these vortex structures. Normally, due to the small distance between them, we cannot resolve the differences in the gyrotropic field produced by the movement of vortices and antivortices and the nucleation of the V-AV pair cannot be fully understood from this type of analysis. For that, we calculated the $\textbf{h}_{g}^{xy}$ component during the V$^-$-AV$^-$ nucleation and separation, which are fundamental steps in the process of core magnetization reversal. Fig.~\ref{Fig4}a shows $\textbf{h}_{g}^{xy}$ for an instant just before the V$^-$-AV$^-$ nucleation and separation. The color map and the arrows illustrate the $z$ magnetization component and  $\textbf{h}_{g}^{xy}$, respectively. The red circle is the original V$^+$ and the blue one is the negative peak.  We can see from Figs.~\ref{Fig4}a-b that $\textbf{h}_{g}^{xy}$ is responsible for the transformation of this wide negative peak to a V$^-$-AV$^-$ pair and acts as a driven force pushing V$^-$. As can be seen Fig.~\ref{Fig4}b, $\textbf{h}_{g}^{xy}$ pushes the AV$^-$ in the direction of V$^+$, producing the pair annihilation. In Fig.~\ref{Fig4}c it is also possible to observe the spin waves generated by the AV$^-$-V$^+$ annihilation. In the last snapshot, we can see the remaining  V$^-$(see Fig.~\ref{Fig4}d). We would like to stress that to our knowledge, this is the first analysis that shows explicitly the dynamics responsible for both the AV$^-$-AV$^-$ nucleation and separation and V$^-$-V$^+$ annihilation. For instance,  considering only the perpendicular component of $\textbf{h}_g$, together with the vorticity conservation, one is able to explain the AV$^-$-V$^+$ annihilation but not the V$^-$-AV$^-$ nucleation and separation process.

In conclusion, we presented a detailed analysis of the influence of $K_z$ on static and dynamic properties of magnetic vortices in disks. We showed that increasing values of $K_z$ produce a growth of the vortex core. In addition, high values of $K_z$ cause a  change in the magnetization pattern. We then showed, by means of dynamical calculations using in-plane magnetic field pulses, that both in-plane and perpendicular components of the gyrotropic field, $\textbf{h}_{g}^{z}$ and $\textbf{h}_{g}^{xy}$, contribute and are fundamental to the understanding of the vortex core magnetization reversal process. 

We thank Fl\'avio Garcia for useful discussions and the Brazilian agencies CNPq and FAPERJ for financial support.


\begin{thebibliography}{10}
\bibitem{Yamada2007}
K. Yamada, S. Kasai, Y. Nakatani, K. Kobayashi, H. Kohno, A.
Thiaville, and T. Ono,
Nature Materials, {\bf 6}, 269 (2007).

\bibitem{Parkin2008}
S.S.P. Parkin, M. Hayashi, and L. Thomas,
Science, {\bf 320}, 190 (2008), and References therein.

\bibitem{Wachowiak2002}
T. Shinjo, T. Okuno, R. Hassdorf, K. Shigeto and T.
Ono, {Science} {\bf 289}, 930 (2000); A. Wachowiak J. Wiebe, M. Bode, O. Pietzsch, M. Morgenstern,
and R. Wiesendanger, Science {\bf 298}, 577 (2002).

\bibitem{Hertel2007}
R. Hertel, S. Gliga, M. F\"ahnle, and C. M. Schneider,
Phys. Rev. Lett. {\bf 98}, 117201 (2007).

\bibitem{Waeyenberge2006}
B. Van Waeyenberge,  A. Puzic, H. Stoll, K. W. Chou, T. Tyliszczak, R.
Hertel, M. Fhnle, H. Brckl, K. Rott, and G. Reiss,
Nature {\bf 444}, 461 (2006).

\bibitem{Kim2007}
K.-S Lee, K.Y. Guslienko, J.-Y Lee, and S.-K. Kim,
Phys. Rev. B {\bf 76}, 174410 (2007).

\bibitem{Eames2002}
P. Eames and E. Dan Dahlberg,
J. Appl. Phys. {\bf 91}, 7986 (2002).

\bibitem{Moutafis2006}
C. Moutafis, S. Komineas, C. A. F. Vaz, J. A. C. Bland, and P. Eames,
Phys. Rev. B {\bf 74}, 214406 (2006).

\bibitem{Fidler}
J. Fidler, and T. Schrefl,
J. Phys. D: Appl. Phys. {\bf 33}, 135 (2000).

\bibitem{Hehn1996}
M. Hehn {\it et al.}, Science  {\bf 272}, 1782 (1996).

\bibitem{Kim2008}
K.Y. Guslienko, Ki-Suk Lee, and Sang-Koog Kim,
Phys. Rev. Lett. {\bf 100}, 027203 (2008).

\bibitem{Xiao2006}
Q.F. Xiao, J. Rudge, B. C. Choi, Y. K. Hong, and G. Donohoe,
Appl. Phys. Lett. {\bf 89}, 262507 (2006).

\bibitem{Thiele1973}
A.A. Thiele,
Phys. Rev. Lett. {\bf 30}, 230 (1973).


\end{thebibliography}
\end{document}